\documentclass[notitlepage,nofootinbib,aps,12pt,tightenlines,%
  showkeys]{revtex4-1}

\pdfoutput=1

\usepackage{amsmath}
\usepackage{amssymb,amsfonts}
\usepackage{bm}
\usepackage[mathcal]{euscript}
\usepackage{graphicx}
\usepackage{subfigure}
\usepackage{hyperref}

\graphicspath{{figs/}{figs_lo/}}

\newcommand{\ie}{\textit{i.e.}}
\newcommand{\etal}{\textit{et~al.}}

\newcommand{\mathnotation}[2]{\newcommand{#1}{\ensuremath{#2}}}

\newcommand{\pow}[2]{#1^#2}

\renewcommand{\l}{\left}			
\renewcommand{\r}{\right}			
\mathnotation{\pd}{\partial}			
\mathnotation{\ldef}{\mathrel{\raisebox{.069ex}{:}\!\!=}}
\mathnotation{\rdef}{\mathrel{=\!\!\raisebox{.069ex}{:}}}
\newcommand{\Order}[1]{\ensuremath{\mathcal{O}\!\l(#1\r)}}

\mathnotation{\grad}{\nabla}			
\mathnotation{\lapl}{\nabla^2}			

\renewcommand{\time}{\tau}			
\mathnotation{\xc}{x}				
\mathnotation{\xv}{\bm{x}}			
\mathnotation{\yc}{y}				
\mathnotation{\uc}{u}				
\mathnotation{\uv}{{\bm{u}}}			
\newcommand{\ev}[1]{\bm{e}_#1}			
\newcommand{\evup}[1]{\bm{e}^#1}		
\newcommand{\euv}[1]{\hat{\bm{e}}_#1}		
\newcommand{\evt}[1]{\tilde{\bm{e}}_#1}		
\newcommand{\evtup}[1]{\tilde{\bm{e}}^#1}	

\mathnotation{\grav}{g}				
\mathnotation{\gravsc}{{\grav_{\mathrm{s}}}}
\mathnotation{\gravv}{\bm{\grav}}		
\mathnotation{\pres}{p}				
\mathnotation{\Hconst}{H}			

\mathnotation{\Metricc}{\mathbb{G}}		
\mathnotation{\metricc}{\mathfrak{g}}		
\mathnotation{\Xc}{X}				
\mathnotation{\Xv}{\bm{\Xc}}			
\mathnotation{\rc}{r}				
\mathnotation{\rv}{\bm{\rc}}			

\mathnotation{\Gcurv}{\mathcal{G}}		

\mathnotation{\eps}{\varepsilon}

\mathnotation{\fmg}{f}				
\mathnotation{\pmg}{p}				
\mathnotation{\qmg}{q}				
\mathnotation{\rmg}{r}				
\mathnotation{\smg}{s}				
\mathnotation{\tmg}{t}				
\mathnotation{\wmg}{w}				

\begin{document}

%
%
%

\title{Chaotic Geodesics}
\thanks{\href{http://dx.doi.org/10.1142/9789812818805_0003}{Proceedings of the conference on
\emph{Chaos, Complexity, and Transport:
Theory and Applications}} (Le Pharo, Marseille, June 2007).}
\author{Jean-Luc Thiffeault}
\affiliation{Department of Mathematics, University of Wisconsin,
  Madison, WI 53706, USA}
\email{jeanluc@math.wisc.edu}
\author{Khalid Kamhawi}
\affiliation{Department of Mathematics, Imperial College London,
  London, SW7 2AZ, UK}

\begin{abstract}
  When a shallow layer of inviscid fluid flows over a substrate, the
  fluid particle trajectories are, to leading order in the layer
  thickness, geodesics on the two-dimensional curved space of the
  substrate.  Since the two-dimensional geodesic equation is a two
  degree-of-freedom autonomous Hamiltonian system, it can exhibit
  chaos, depending on the shape of the substrate.  We find chaotic
  behaviour for a range of substrates.
\end{abstract}


\maketitle

\section{Introduction}

Many well-known physical systems take the form of geodesic flow on a
manifold.  For instance, Euler's equation can be though of as a
geodesic flow in the space of volume-preserving
diffeomorphisms,~\cite{Arnold1966a,Arnold1969b,Arnold,MarsdenRatiu,%
  ArnoldTopo,Watanabe2007} and free rigid body motion as geodesic flow
in~$SO(3)$.  In both cases, the metric on the space corresponds to the
kinetic energy norm.  It is also known that the \emph{geodesic
  deviation equation}~\cite{Wald} describes the stability of such
flows.  For instance, a space of negative curvature will lead to
divergence of trajectories, and hence to chaos if the space is compact.
But compact spaces of strictly negative curvature are hard to come by
in the real world, to say the least.  If we expect the negative
curvature to lead to chaotic geodesics, we are better off looking for
spaces with non-sign-definite curvature, but such that the averaging
of the curvature over trajectories leads to chaos (\ie, the negative
curvature `wins').

In this contribution we will discuss a system which is
physically-motivated and leads to chaotic geodesics.  This system is
the flow of a shallow layer of ideal, irrotational fluid on a curved
substrate.  Following Rienstra~\cite{Rienstra1996}, we will show
that, to leading order, the governing equation can be solved in terms
of characteristics.  Moreover, the characteristics are geodesics on
the curved substrate, possibly modified by gravity if it is present.

\begin{figure}
\begin{center}
\includegraphics[width=.7\textwidth]{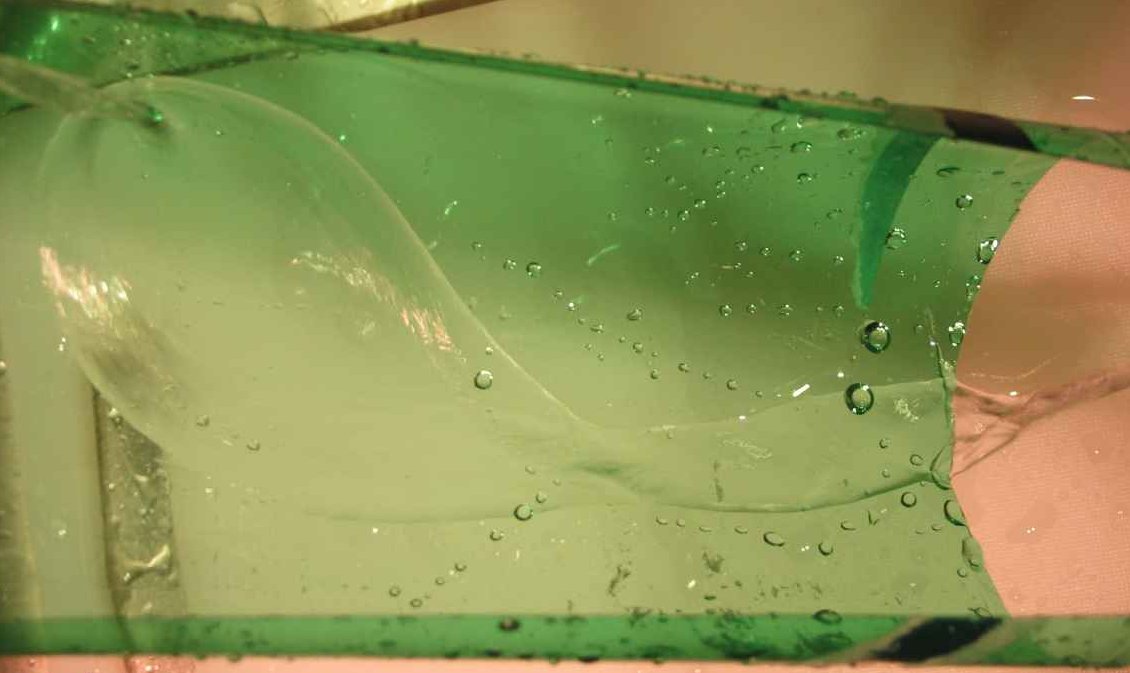}

\vspace*{.6cm}

\includegraphics[width=.7\textwidth]{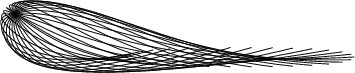}

\vspace*{.6cm}

\includegraphics[width=.7\textwidth]{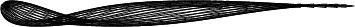}
\end{center}

\caption{Experiment in the kitchen sink, using a cut-open plastic bottle
  (top).  The jet from the faucet impacts the inclined bottle.  The pattern is
  qualitatively well reproduced by following fluid trajectories emanating from
  a point source on a cylinder (middle).  However, if we pursue the
  trajectories further (bottom), the ideal theory presented here predicts that
  the flow should crawl back up the side to the same initial height.  The
  discrepancy is clearly due to dissipation.}
\label{fig:cylinder}
\end{figure}

Of course, the chaotic trajectories have a nasty tendency to cross and
form caustics everywhere.  Hydrodynamically, caustics are usually
manifested as hydraulic jumps or so-called `mass
tubes,'~\cite{Edwards2008} visible as a thicker edge region of the
fluid in Fig.~\ref{fig:cylinder} (top).
Edwards~\etal~\cite{Edwards2008} have recently described these mass
tubes using the theory of
`delta-shocks.'~\cite{Bouchut1994,LiZhangYang,Yang1999,Li2001,Edwards2008}
At this point, we are unable to apply this theory to our problem,
which means that solutions become dubious after characteristics begin
to cross.  Unfortunately, since our characteristics are chaotic, they
tend to cross a lot.  Nevertheless, we believe that studying the basic
properties of this geodesic flow is worthwhile as a first stab at
describing the transport properties of flows on curved substrates.  In
addition, the geodesic flow we present is an interesting mathematical
system, with rich dynamics that deserve to be studied on their own.

Having gotten these disclaimers out of the way, let us proceed with
the analysis.  We hall do this in stages.  In Section~\ref{sec:coords}
we introduce a curved non-orthogonal coordinate system to describe the
substrate, singling out a direction normal to the substrate.  In
Section~\ref{sec:dynamics} we use this direction to expand our fluid
equations and derive a shallow-layer form.  We show that the resulting
equation can be solved in terms of characteristics, which are
geodesics on the two-dimensional curved substrate, modified by
gravity.  In Section~\ref{sec:advection} we look at specific numerical
solutions for particle trajectories, and in Section~\ref{sec:Lyapunov}
we speculate on their chaotic nature.  We offer some closing comments
in Section~\ref{sec:discussion}.

\section{Coordinate System}
\label{sec:coords}

\subsection{Separating the Shallow Direction}
\label{sec:thindir}

In our problem, fluid motion occurs over a curved substrate of arbitrary
shape.  The direction normal to the substrate is special in that it defines
the direction in which the fluid layer is assumed `shallow.'  Hence, it is
convenient to locate a point~$\rv$ in the fluid as
\begin{equation}
  \rv(\xc^1,\xc^2,\yc) = \Xv(\xc^1,\xc^2) + \yc\,\euv{3}(\xc^1,\xc^2)
  \label{eq:r}
\end{equation}
where~$\Xv(\xc^1,\xc^2)$ is the location of the substrate,~$\euv{3}$ is a unit
vector normal to the substrate, and~$\yc$ is the perpendicular distance
from~$\rv$ to the substrate.  The coordinates~$\xc^1$ and~$\xc^2$ are
substrate coordinates used to localise points on the substrate.  For example,
in Section~\ref{sec:monge} we will use the Monge
parametrisation,~$\Xv=\begin{pmatrix}\xc^1 & \xc^2 &
\fmg(\xc^1,\xc^2)\end{pmatrix}^T$, where~$\fmg$ gives the height of the
substrate.

The tangent vectors to the substrate are
\begin{equation}
  \ev{\alpha} \ldef \pd_\alpha\Xv
  \label{eq:tangsubstr}
\end{equation}
where~$\pd_\alpha \ldef \pd/\pd\xc^\alpha$.  The coordinate vectors
associated with the coordinate system are found from~\eqref{eq:r},
\begin{equation}
  \evt{\alpha} \ldef \pd_\alpha\rv = \ev{\alpha} + \Order{\yc}\,,\qquad
  \euv{3} \ldef \frac{\pd\rv}{\pd \yc}\,.
  \label{eq:tang}
\end{equation}
where Greek indices only take the value~$1$ or~$2$.  Note that
the~$\ev{\alpha}$ are not necessarily orthogonal or normalised.

We adopt the convention that quantities with a tilde are evaluated in the
`bulk' (away from the substrate), and thus depend on~$\yc$, whilst those
without the tilde are `substrate' quantities and do not depend on~$\yc$.
Thus,~$\evt{\alpha}(\xc^1,\xc^2,0)=\ev{\alpha}(\xc^1,\xc^2)$.  The
three-dimensional metric tensor~$\tilde\metricc_{ab}$ has components
\begin{equation*}
  \tilde\metricc_{\alpha\beta} \ldef \evt{\alpha}\cdot\evt{\beta} =
  \widetilde\Metricc_{\alpha\beta}\,,\quad
  \tilde\metricc_{\alpha3} \ldef \evt{\alpha}\cdot\euv{3} = 0\,,\quad
  \tilde\metricc_{33} \ldef \euv{3}\cdot\euv{3} = 1\,,
\end{equation*}
where
\begin{equation}
\begin{split}
  \widetilde\Metricc_{\alpha\beta} &\ldef \evt{\alpha}\cdot\evt{\beta} =
  \Metricc_{\alpha\beta} + \Order{\yc}\,,\\
  \Metricc_{\alpha\beta} &\ldef \ev{\alpha}\cdot\ev{\beta}\,.
\end{split}
\label{eq:GtG}
\end{equation}
The three-dimensional metric tensor is thus block-diagonal, and the~$\yc$
coordinate is unstretched compared to the Cartesian coordinate system.  It
measures the true perpendicular distance from the substrate to a point in the
fluid.

Given the substrate vectors~$\ev{\alpha}$, it is easy to solve for the
covectors~$\evup{\alpha}$, which are such
that \hbox{$\evup{\alpha}\cdot\ev{\beta}={\delta_\beta}^\alpha$}.  Then the
bulk covectors are
\begin{equation}
  \evtup{\alpha} = \evup{\alpha} + \Order{\yc},
  \label{eq:cotang}
\end{equation}
to leading order in~$\yc$.  From~\eqref{eq:cotang}, we find the
inverse metrics,
\begin{equation}
\begin{split}
  \widetilde\Metricc^{\alpha\beta} &= \evtup{\alpha}\cdot\evtup{\beta}
  = \Metricc^{\alpha\beta} + \Order{\yc},\label{eq:Metrictinv}\\
  \Metricc^{\alpha\beta} &= \evup{\alpha}\cdot\evup{\beta}\,.
\end{split}
\end{equation}
Higher-order terms in~$\yc$ will not be needed.

\subsection{Substrate Coordinates}
\label{sec:monge}

For most applications in the literature of thin films and shallow
layers, orthonormal coordinates have been the coordinates of choice.
This is because the main substrate shapes that have been treated are
planes, cylinders, and spheres, where orthonormal coordinates are
readily available.  For a general substrate shape, orthonormal
coordinates are difficult to construct and require numerical
integration.  Singularities (umbilics) also cause
problems.~\cite{Kreyszig} For our application---flow down a curved
substrate---the Monge representation of a surface~\cite{Flanders} is
the most convenient.

The Monge representation is a glorified name for a parametrisation of the
substrate by
\begin{equation}
  \Xv(\xc^1,\xc^2)
  =\begin{pmatrix}\xc^1 & \xc^2 & \fmg(\xc^1,\xc^2)\end{pmatrix}^T
  \label{eq:Xsurf}
\end{equation}
in three-dimensional Cartesian space.  Following standard
notation,~\cite{Flanders} we define
\begin{subequations}
\begin{gather}
  \pmg = \pd_1\fmg,\quad
  \qmg = \pd_2\fmg,\quad\\
  \rmg = \pd_1\pd_1\fmg,\quad
  \smg = \pd_1\pd_2\fmg,\quad
  \tmg = \pd_2\pd_2\fmg.
\end{gather}%
\label{eq:mongeparams}%
\end{subequations}%
The unnormalised, nonorthogonal tangents~$\ev{1}$ and~$\ev{2}$ are
\begin{equation*}
  \ev{1} = \pd_1\Xv
  = \begin{pmatrix}1 & 0 & \pmg\end{pmatrix}^T,\qquad
  \ev{2} = \pd_2\Xv
  = \begin{pmatrix}0 & 1 & \qmg\end{pmatrix}^T,
\end{equation*}
and their normalised cross product gives the normal to the substrate,
\begin{equation*}
  \euv{3} = \frac{1}{\wmg}\,\begin{pmatrix}-\pmg & -\qmg & 1\end{pmatrix}^T.
\end{equation*}
The corresponding covectors are
\begin{equation*}
  \evup{1} = \frac{1}{\pow{\wmg}{2}}\begin{pmatrix}
    (1+\pow{\qmg}{2}) & -\pmg\qmg & \pmg\end{pmatrix}^T,\qquad
  \evup{2} = \frac{1}{\pow{\wmg}{2}}\begin{pmatrix}-\pmg\qmg
    & (1+\pow{\pmg}{2}) & \qmg\end{pmatrix}^T
\end{equation*}
and~$\euv{3}$ is its own covector.  The metric tensor of the substrate and its
inverse are
\begin{subequations}
\begin{align}
  \{\Metricc_{\alpha\beta}\} &= \ev{\alpha}\cdot\ev{\beta}
  = \phantom{\frac{1}{\pow{\wmg}{2}}}
  \begin{pmatrix}1+\pow{\pmg}{2} & \pmg\qmg \\ \pmg\qmg &
  1+\pow{\qmg}{2}\end{pmatrix},\\
    \quad
  \{\Metricc^{\alpha\beta}\} &= \evup{\alpha}\cdot\evup{\beta}
  = \frac{1}{\pow{\wmg}{2}}
  \begin{pmatrix}1+\pow{\qmg}{2} & -\pmg\qmg \\ -\pmg\qmg &
  1+\pow{\pmg}{2}\end{pmatrix},
\end{align}%
\label{eq:GMonge}%
\end{subequations}%
with determinant
\begin{equation*}
  \wmg = \bigl(\det\,\Metricc_{\alpha\beta}\bigr)^{1/2}
  = \sqrt{1 + \pow{\pmg}{2} + \pow{\qmg}{2}}\,.
\end{equation*}
Finally, we will need the Christoffel
symbols~$\Gamma^\sigma_{\alpha\beta}$, defined by
\begin{equation}
  \Gamma^\sigma_{\alpha\beta} = \tfrac{1}{2}\Metricc^{\sigma\gamma}\l(
  \pd_\alpha\Metricc_{\gamma\beta}
  + \pd_\beta\Metricc_{\gamma\alpha}
  - \pd_\gamma\Metricc_{\alpha\beta}\r)
  \label{eq:Christoffel}
\end{equation}
and in Monge coordinates given by
\begin{equation}
  \Gamma^1_{\alpha\beta}
  = \frac{\pmg}{\wmg^2}
  \begin{pmatrix}\rmg & \smg \\ \smg & \tmg \end{pmatrix},\qquad
  \Gamma^2_{\alpha\beta}
  = \frac{\qmg}{\wmg^2}
  \begin{pmatrix}\rmg & \smg \\ \smg & \tmg \end{pmatrix}.
  \label{eq:ChristoffelMonge}
\end{equation}
The Christoffel symbols arise when taking covariant
derivatives.~\cite{Wald,Synge,Schutz} Note that in \eqref{eq:Christoffel} we
used the usual convention that repeated indices are summed.

We write the normalised gravity vector as
\begin{equation}
  \gravv = (\sin\theta\cos\phi\ \ \sin\theta\sin\phi\ \ -\cos\theta)^T\,,
  \label{eq:gravvec1}
\end{equation}
so that the inclination angle~$\theta$
is zero for gravity pointing downwards, and for~$\phi\in(-\pi/2,\pi/2)$
positive~$\theta$ induces flow in the positive~$\xc^1$ direction.  Then we
have the components
\begin{equation}
\begin{split}
\gravsc^1 &= \gravv\cdot\evup{1}
  = -\l(\pmg\,\cos\theta + \pmg\qmg\,\sin\theta\sin\phi
  - (1+\pow{\qmg}{2})\sin\theta\cos\phi\r)/\pow{\wmg}{2}\,,\\
\gravsc^2 &= \gravv\cdot\evup{2}
  = -\l(\qmg\,\cos\theta + \pmg\qmg\,\sin\theta\cos\phi
  - (1+\pow{\pmg}{2})\sin\theta\sin\phi\r)/\pow{\wmg}{2}\,,
\end{split}%
\label{eq:gravvec}%
\end{equation}%
The specific parametrisation of the substrate introduced in this
section will not be needed in the derivation of the equations of
motion (Section~\ref{sec:dynamics}), only in their solution.  Hence, a
different parametrisation could be used if called for by the geometry
of the substrate.  For instance, flow down a curved filament is better
parametrised by cylindrical coordinates, or if the substrate has
overhangs (making~$\fmg$ multivalued) coordinates based on arc length
are preferable.

\section{Equations of Motion}
\label{sec:dynamics}

Now that we've set up an appropriate coordinate system on our curved
substrate, we need some dynamical equations of motion for the fluid.
We assume an inviscid, irrotational fluid with a free surface
at~$\yc=\eta(\xc^1,\xc^2)$, with slip boundary conditions at the
substrate~$\yc=0$.  The pressure on the free surface is assumed
constant (zero).  We also assume the flow is steady and irrotational,
so that the the velocity can be written in terms of a scalar
potential,~$\uv=\grad\varphi$.  The equations satisfied by the fluid
are then
\begin{subequations}
\begin{alignat}{2}
  \lapl\varphi &= 0, \qquad && \text{mass conservation;}
  \label{eq:masscons} \\
  \tfrac{1}{2}\l\lvert\grad\varphi\r\rvert^2 + \frac{\pres}{\rho}
  - \gravv\cdot\rv &= \Hconst, \qquad && \text{Bernoulli's law;}
  \label{eq:Bernoulli}
\end{alignat}
\end{subequations}
where~$\Hconst$ is a constant, with boundary conditions
\begin{subequations}
\begin{alignat}{2}
  \pd_\yc\varphi &= 0 \quad &\text{at } \yc = 0,\qquad
  &\text{no-throughflow at substrate;}
  \label{eq:slip}\\
  \grad\varphi\cdot\grad\eta &= \pd_\yc\varphi\quad
  &\text{at } \yc = \eta, \qquad
  &\text{kinematic condition at free surface;}\label{eq:kinem}\\
  \pres &= 0 \quad
  &\text{at } \yc = \eta, \qquad
  &\text{constant pressure at free surface}.
  \label{eq:pres}
\end{alignat}\label{eq:BCs}%
\end{subequations}%
In terms of our curvilinear coordinates,
equation~\eqref{eq:Bernoulli} becomes
\begin{equation}
  \widetilde\Metricc^{\alpha\beta}\,\pd_\alpha\varphi\,\pd_\beta\varphi
  + (\pd_\yc\varphi)^2 + \frac{2\pres}{\rho}
  - 2\gravv\cdot\rv = 2\Hconst.
\end{equation}

\subsection{Small-parameter Expansion}

Now we assume that the fluid layer is shallow, so that~$\yc$ is
proportional to~$\eps$.  After replacing~$\yc$ by~$\eps\yc$,
Eq.~(\ref{eq:Bernoulli}) becomes
\begin{equation}
  \l(\Metricc^{\alpha\beta}
  + \Order{\eps}\r)\pd_\alpha\varphi\,\pd_\beta\varphi
  + \eps^{-2}(\pd_\yc\varphi)^2 + \frac{2\pres}{\rho}
  - 2\gravv\cdot(\Xv +\eps\,\yc\,\euv{3})= 2\Hconst.
  \label{eq:Bernoulliexp}
\end{equation}
We also expand~$\varphi$ in powers of~$\eps$,
\begin{equation}
  \varphi(\xc^1,\xc^2,\yc) = \varphi_{(0)} + \eps\,\varphi_{(1)}
  + \eps^2\,\varphi_{(2)} + \dots.
\end{equation}
The leading-order term in~\eqref{eq:Bernoulliexp} occurs at
order~$\eps^{-2}$, and gives~$\pd_\yc\varphi_{(0)}=0$.  Hence, we
have~$\varphi_{(0)}=\Phi(\xc^1,\xc^2)$ independent of~$\yc$.  The next
nontrivial terms are at order~$\eps^0$,
\begin{equation}
  \Metricc^{\alpha\beta}\pd_\alpha\Phi\,\pd_\beta\Phi
  + (\pd_\yc\varphi_{(1)})^2 + \frac{2\pres}{\rho}
  - 2\gravv\cdot\Xv = 2\Hconst.
  \label{eq:woohoo}
\end{equation}
We evaluate the whole of~\eqref{eq:woohoo} at~$\yc=\eta$, and use the boundary
conditions~\eqref{eq:kinem} (expanded in~$\eps$, $\pd_\yc\varphi_{(1)}=0$ at
$\yc=\eta$) and~\eqref{eq:pres}, to obtain
\begin{equation}
  \Metricc^{\alpha\beta}\pd_\alpha\Phi\,\pd_\beta\Phi
  - 2\gravv\cdot\Xv = 2\Hconst.
  \label{eq:woohoo2}
\end{equation}
This is the equation that we need to solve to find the leading-order
velocity potential~$\Phi(\xc^1,\xc^2)$.  We discuss the method of
solution in the next section.  Note that we will not need to solve the
mass conservation equation~(\ref{eq:masscons}): at leading order, it
only serves to find the fluid height once the velocity field is
obtained.

\begin{figure}
\begin{center}
\includegraphics[width=.31\textwidth]{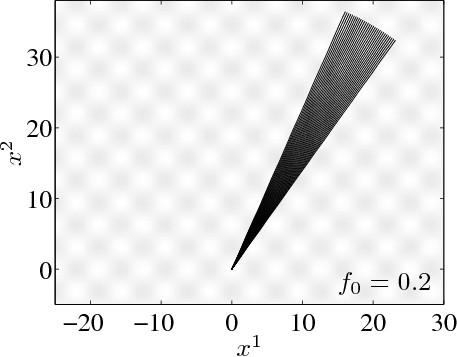}
\hspace{.5em}
\includegraphics[width=.31\textwidth]{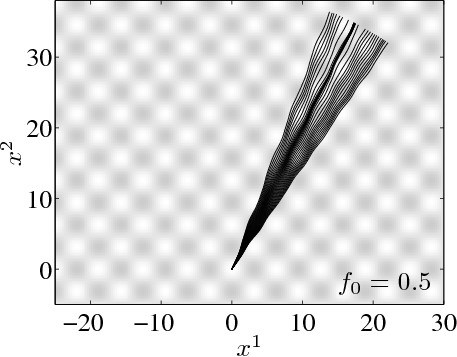}
\hspace{.5em}
\includegraphics[width=.31\textwidth]{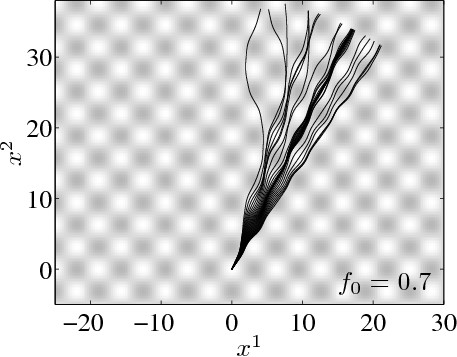}

\vspace*{.5em}

\includegraphics[width=.31\textwidth]{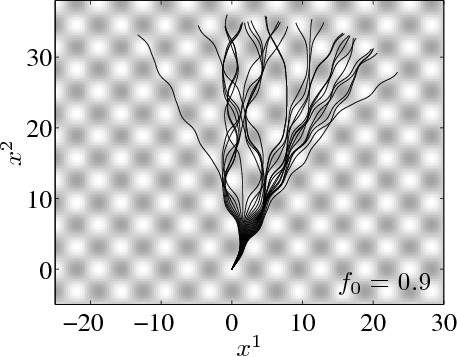}
\hspace{.5em}
\includegraphics[width=.31\textwidth]{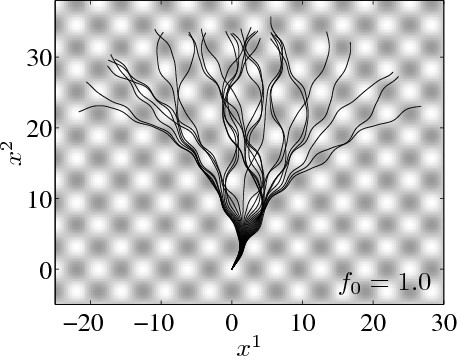}
\hspace{.5em}
\includegraphics[width=.31\textwidth]{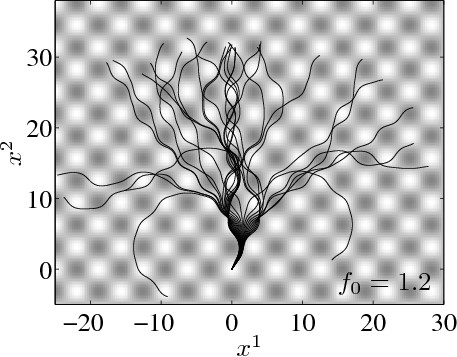}
\end{center}

\caption{A pencil of 30 trajectories starting from the origin at
  different angles, each with initial kinetic energy~$1/2$.  The
  substrate has shape~$\fmg(\xc^1,\xc^2) =
  \fmg_0\,\cos\xc^1\cos\xc^2$, and the plots are for different values
  of~$\fmg_0$.  Gravity is turned off:~$\grav=0$.  The grey background
  shows the periodicity of the substrate.}
\label{fig:nograv}
\end{figure}

\subsection{Solution in Terms of Characteristics}

As pointed out by Rienstra,~\cite{Rienstra1996} the trick to solving
Eq.~\eqref{eq:woohoo2} is to use the method of characteristics.
To do this, we differentiate~\eqref{eq:woohoo2} with respect
to~$\xc^\gamma$, which gets rid of the constant~$\Hconst$,
\begin{equation}
  2\Metricc^{\alpha\beta}\pd_\alpha\Phi\,\pd_\gamma\pd_\beta\Phi
  + \pd_\gamma\Metricc^{\alpha\beta}\pd_\alpha\Phi\,\pd_\beta\Phi
  = 2\gravv\cdot\pd_\gamma\Xv.
  \label{eq:gammader}
\end{equation}
The horizontal components of velocity are~$\dot\xc^\alpha =
\Metricc^{\alpha\beta}\pd_\beta\Phi$, where the overdot denotes a time
derivative; hence,
\begin{equation}
  \pd_\gamma\pd_\beta\Phi = \pd_\beta(\Metricc_{\gamma\delta}\dot\xc^\delta)
  = \Metricc_{\gamma\delta}\pd_\beta\dot\xc^\delta
  + \pd_\beta\Metricc_{\gamma\delta}\dot\xc^\delta\,.
  \label{eq:Phi2}
\end{equation}
From the chain rule, we have~$\ddot\xc^\delta =
\pd_\beta\dot\xc^\delta\dot\xc^\beta$; using this and~\eqref{eq:Phi2}
in~\eqref{eq:gammader}, we find, after dividing by two,
\begin{equation}
  \Metricc_{\gamma\delta}\ddot\xc^\delta
  + \pd_\beta\Metricc_{\gamma\delta}\dot\xc^\beta\dot\xc^\delta
  + \tfrac{1}{2}\pd_\gamma\Metricc^{\alpha\beta}
  \Metricc_{\alpha\delta}\Metricc_{\beta\rho}\,
  \dot\xc^\delta\dot\xc^\rho = \gravv\cdot\ev{\gamma}.
\end{equation}
Now we multiply by~$\Metricc^{\sigma\gamma}$, and obtain after an integration
by parts and a bit of manipulation
\begin{equation}
  \ddot\xc^\sigma
  + \Gamma^\sigma_{\alpha\beta}\,
  \dot\xc^\alpha\dot\xc^\beta = \gravv\cdot\evup{\sigma}
  \label{eq:geodesic}
\end{equation}
where the~$\Gamma^\sigma_{\alpha\beta}$ are defined
by~\eqref{eq:ChristoffelMonge}.  Equation~\eqref{eq:geodesic}
describes \emph{geodesics} in the curved coordinates of the substrate,
under the influence of gravity.  In the absence of gravity, the fluid
trajectories are essentially going in straight lines in the curved
substrate coordinates.  (In general relativity, unlike here, the
gravity determines the curvature of space.)

If we define the \emph{covariant derivative} of a vector~$V^\sigma$
along the trajectory~\cite{Wald,Synge,Schutz,Thiffeault2001e},
\begin{equation}
  \frac{D}{D\time}\,V^\sigma \ldef \dot V^\sigma
  + \Gamma^\sigma_{\alpha\beta}\,\dot\xc^\alpha V^\beta\,,
  \label{eq:covderV}
\end{equation}
where~$\time$ is the time (to avoid confusion with~$\tmg$
in Eq.~(\ref{eq:mongeparams})), then the geodesic
equation~\eqref{eq:geodesic} takes the more intuitive form
\begin{equation}
  \frac{D}{D\time}\dot\xc^\sigma
  = \gravv\cdot\evup{\sigma}
  \label{eq:geodesic2}
\end{equation}
which looks a lot like Newton's second law, but here it incorporates
the constraint that fluid particles remain on the substrate.

Equation~\eqref{eq:geodesic} is a two degree-of-freedom autonomous
Hamiltonian system, with the energy~$\Hconst$ defined by
equation~\eqref{eq:woohoo2} as an invariant.  Hence, any other
invariant will make the system integrable, and rule out chaos.  In
particular, a surface with a translational symmetry cannot exhibit
chaos.


\section{Fluid Particle Trajectories}
\label{sec:advection}

\begin{figure}
\begin{center}
\includegraphics[width=.27\textwidth]{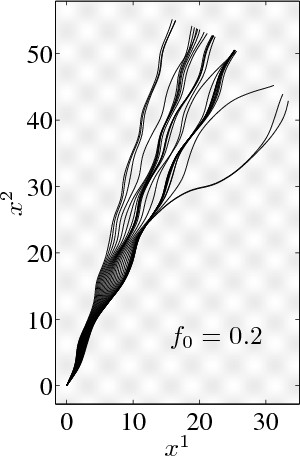}
\hspace{.5em}
\includegraphics[width=.36\textwidth]{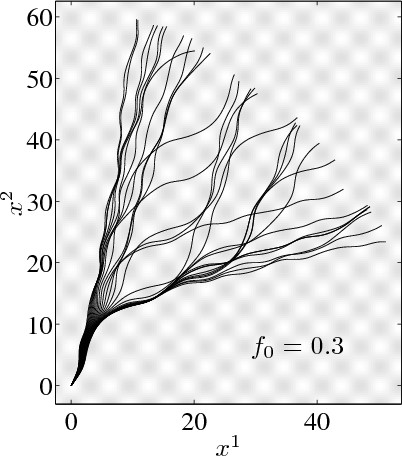}

\vspace*{.5em}

\includegraphics[width=.36\textwidth]{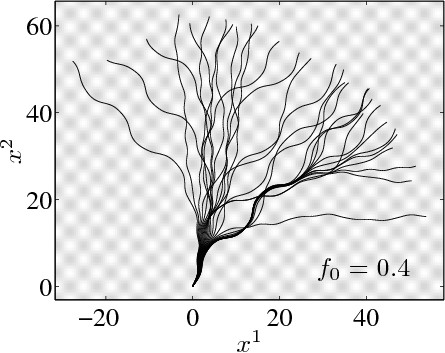}
\hspace{.5em}
\includegraphics[width=.36\textwidth]{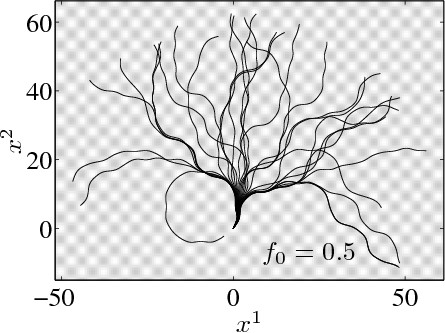}
\end{center}

\caption{Same parameters as in Fig.~\ref{fig:nograv}, but with gravity
  turned on:~$\grav=1$.  The trajectories exhibit chaos-like behaviour
  for much lower substrate height, since they begin at the top of a
  bump and thus have potential energy to draw upon.}
\label{fig:grav}
\end{figure}

\begin{figure}
\begin{center}
\includegraphics[width=.31\textwidth]{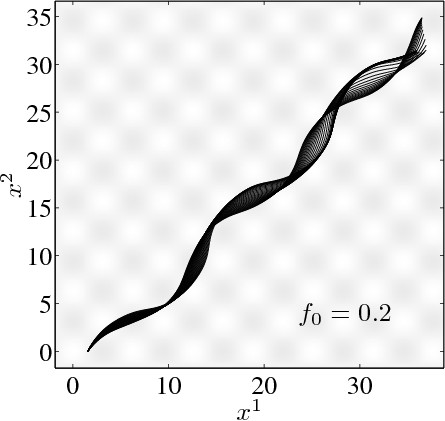}
\hspace{.5em}
\includegraphics[width=.31\textwidth]{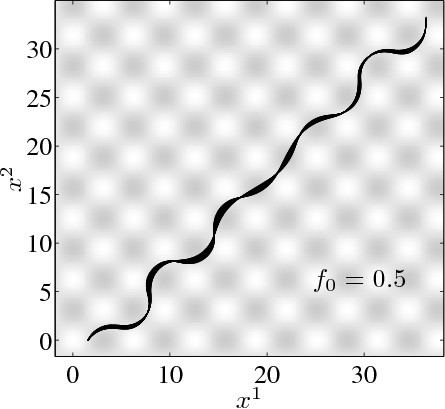}
\hspace{.5em}
\includegraphics[width=.31\textwidth]{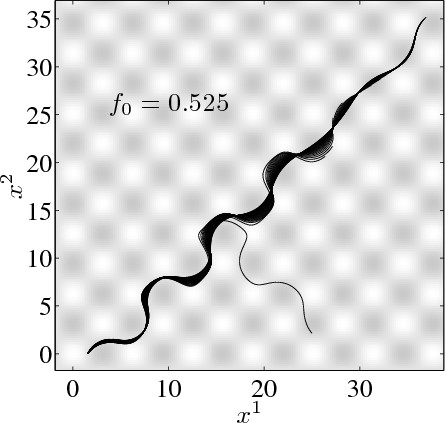}

\vspace*{.5em}

\includegraphics[width=.31\textwidth]{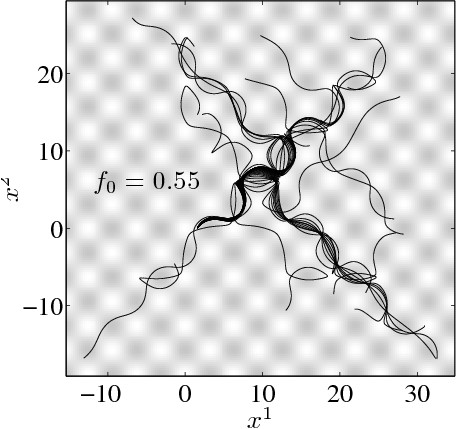}
\hspace{.5em}
\includegraphics[width=.31\textwidth]{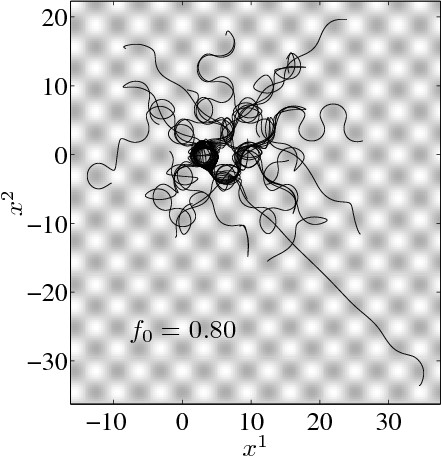}
\hspace{.5em}
\includegraphics[width=.31\textwidth]{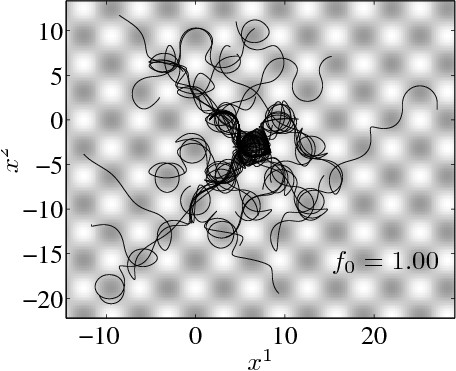}
\end{center}

\caption{Same parameters as in Fig.~\ref{fig:grav}, but with initial
  condition~$(\xc^1,\xc^2)=(\pi/2,0)$.  For larger substrate
  amplitudes, the system is dominated by `rimming,' where particles
  skim a depression before moving to the next one, or sometimes
  undergo long flights.}
\label{fig:gravpi2}
\end{figure}

To get a feel for the possible range of behaviour of fluid
trajectories, we now solve the geodesic equation~\eqref{eq:geodesic}
for a range of substrates.  We shall always use a set of fluid
trajectories starting at the same spatial point, with the same initial
kinetic energy but different direction.  This models a point source,
or a thin jet impacting the substrate.

First, following Rienstra~\cite{Rienstra1996} we solve the equations
on a cylindrical substrate.  Figure~\ref{fig:cylinder} (middle) shows
some trajectories, all emanating from the same point.  Qualitatively,
the pattern captures well the observed behaviour of a jet (from a
faucet) impacting the inside of a cut-out plastic bottle
(Fig.~\ref{fig:cylinder}, top).  However, if we pursue the
trajectories further (Fig.~\ref{fig:cylinder}, bottom), we see that
they crawl back up the side of the cylinder, with no loss of energy,
in contrast to the experimental picture.  This comes from neglecting
the hydraulic jumps that occur, as well as
viscosity.~\cite{Edwards2008}  Observe that the trajectories follow a
very ordered pattern, and are definitely not chaotic.  This is as
expected, since there is a symmetry direction, and so the motion is
integrable (Section~\ref{sec:dynamics}).

Next we move on to more complex substrates.  Since there is basically
an infinity of choices here, we limit ourselves to periodic substrates
with shape
\begin{equation}
  \fmg(\xc^1,\xc^2) = \fmg_0\,\cos\xc^1\cos\xc^2
\end{equation}
for various values of~$\fmg_0$.  The other variables in the system are
the strength of gravity (which can be chosen as unity if it is not
zero, by rescaling the substrate height) and its orientation (as given
by the angles~$\theta$ and~$\phi$ in Eq.~\eqref{eq:gravvec1}).

\begin{figure}
\begin{center}
  \includegraphics[width=.5\textwidth]{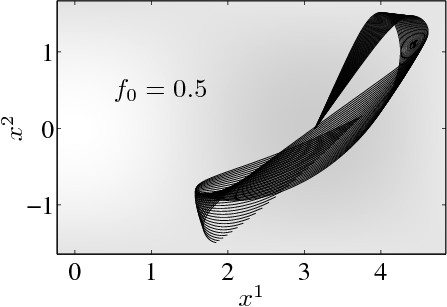}
\end{center}

\caption{Same parameters as in Fig.~\ref{fig:grav}, but with initial
    condition~$(\xc^1,\xc^2)=(\pi,0)$.  The particles begin at the
    bottom of the potential well, and they do not have the energy to
    escape.}
\label{fig:gravpi}
\end{figure}

Figure~\ref{fig:nograv} shows a pencil of 30 trajectories starting
from the origin at different angles, each with initial kinetic
energy~$1/2$, for different values of~$\fmg_0$, in the absence of
gravity.  The first two cases display regular behaviour, but for
substrate heights~$\fmg_0=0.7$ there is chaotic-like behaviour.  These
are, however, fairly extreme values of~$\fmg_0$, corresponding to
heavily-deformed substrates.  Our expansion should be able to
accommodate this, since the variations in the substrate height are not
assumed small (only those in the fluid thickness are).  For extreme
heights ($\fmg_0=1.2$, last case in Fig.~\ref{fig:nograv}), some
trajectories actually backfire and come around the initial point.

Figure~\ref{fig:grav} shows results for the same parameters as
Fig.~\ref{fig:nograv}, but with gravity~$\grav=1$.  The inclination is
nil ($\theta=0$).  It is clear that chaotic-like behaviour sets in for
much smaller values of~$\fmg_0$, even showing backscatter
for~$\fmg_0=0.5$ in the last frame.  This is because the fluid
elements can now draw on the potential energy they inherit from
starting at the top of the bump.

This suggests that, in the presence of gravity, the results should be
substantially different if we start elsewhere on the substrate.
Figure~\ref{fig:gravpi2} shows simulations with the same parameters as
in Fig.~\ref{fig:grav}, but starting at~$(\xc^1,\xc^2)=(\pi/2,0)$,
some way down the bump.  The motion is then confined to narrow
channels for moderate~$\fmg_0$.  But for larger substrate amplitudes,
the system is dominated by `rimming,' where particles skim a
depression before moving to the next one, or sometimes undergo long
flights.  This is a similar situation to basketball (or golf), where
the ball turns around the hoop (or cup) a while before deciding to go
in or out.  If we take an initial condition at the bottom of the
bump,~$(\xc^1,\xc^2)=(\pi,0)$, then the trajectories do not have
enough energy to escape the potential well (Fig.~\ref{fig:gravpi}).

\begin{figure}
\includegraphics[width=.31\textwidth]{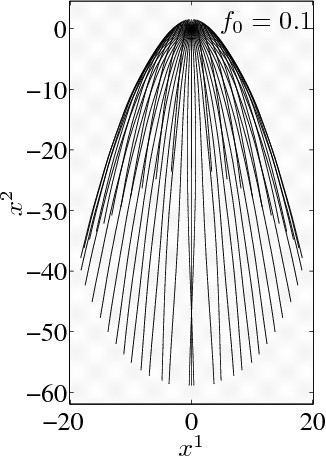}
\hspace{.5em}
\includegraphics[width=.31\textwidth]{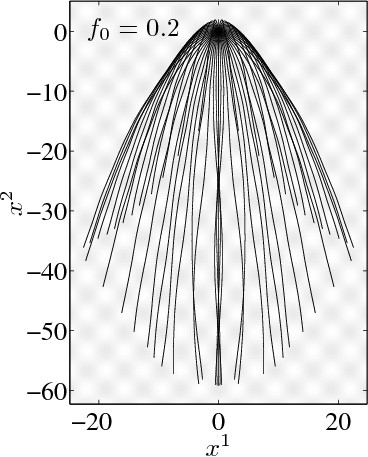}
\hspace{.5em}
\includegraphics[width=.31\textwidth]{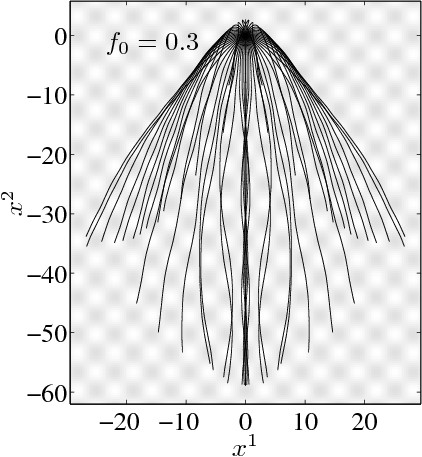}

\vspace*{.5em}

\includegraphics[width=.31\textwidth]{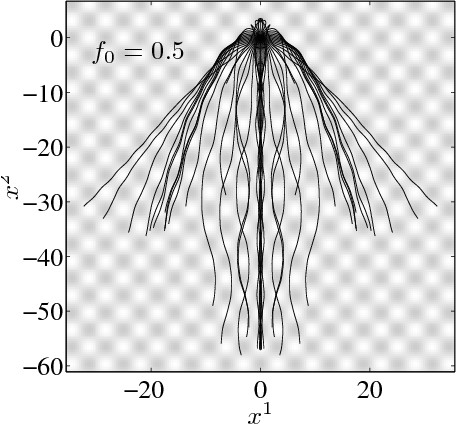}
\hspace{.5em}
\includegraphics[width=.31\textwidth]{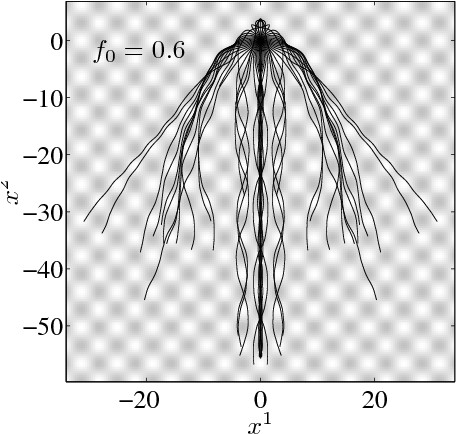}
\hspace{.5em}
\includegraphics[width=.31\textwidth]{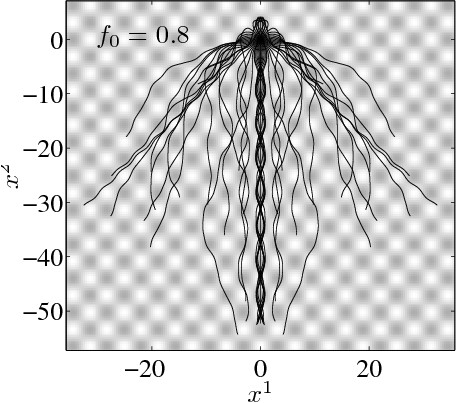}

\caption{Same parameters as in Fig.~\ref{fig:grav}, but with
  incline~$\theta=\pi/8$ and~$\phi=-\pi/2$.  For larger bumps there is
  a `shadow' effect, particularly for~$\fmg_0=0.6$.}
\label{fig:gravinc}
\end{figure}

Finally, in Fig.~\ref{fig:gravinc} we illustrate the effect of inclining
the substrate at an angle~$\theta=\pi/8$.  With~\hbox{$\phi=-\pi/2$}
the trajectories flow `downhill,' in the negative~$\xc^2$ direction,
modified by the bumps.  The system still appears to becomes chaotic
for larger~$\fmg_0$.  Larger bumps induce a `shadow' effect, where
they prevent fluid from flowing behind them (in particular
for~$\fmg_0=0.6$).

\section{Lyapunov Exponents and Chaos}
\label{sec:Lyapunov}

We now investigate whether the behaviour described in the previous
section is chaotic or not.  The first variation of the geodesic
Eq.~\eqref{eq:geodesic} gives
\begin{equation}
  \delta\ddot\xc^\sigma
  + 2\Gamma^\sigma_{\alpha\beta}\,\dot\xc^\alpha\delta\dot\xc^\beta
  + \pd_\gamma\Gamma^\sigma_{\alpha\beta}\,\dot\xc^\alpha\dot\xc^\beta
  \delta\xc^\gamma
  = -\gravv\cdot\evup{\tau}\,\Gamma^\sigma_{\gamma\tau}\delta\xc^\gamma
  \label{eq:dgeodesic}
\end{equation}
where we have used~\hbox{$\pd_\gamma\evup{\sigma} =
  -\Gamma^\sigma_{\gamma\tau}\,\evup{\tau}$}.  This equation can be massaged
into the \emph{geodesic deviation equation},~\cite{Wald,geodesic_deviation}
\begin{equation}
  \frac{D^2}{D\time^2}\delta\xc^\sigma
  +
  {R^\sigma}_{\beta\gamma\alpha}\,\dot\xc^\alpha\dot\xc^\beta\delta\xc^\gamma
  = 0
\end{equation}
where~$D/D\time$ is defined in~\eqref{eq:covderV}, and
\begin{equation}
  {R^\sigma}_{\beta\gamma\alpha} \ldef
  \pd_\gamma\Gamma^\sigma_{\alpha\beta}
  - \pd_\alpha\Gamma^\sigma_{\gamma\beta}
  + \Gamma^\sigma_{\gamma\lambda}\Gamma^\lambda_{\alpha\beta}
  - \Gamma^\sigma_{\alpha\lambda}\Gamma^\lambda_{\gamma\beta}
  \label{eq:Riemanncurv}
\end{equation}
is the Riemann curvature tensor.  For two-dimensional surfaces, the
curvature tensor simplifies to
\begin{equation}
  {R^\sigma}_{\beta\gamma\alpha} = \Gcurv
  \l({\delta^\sigma}_\gamma\,\Metricc_{\beta\alpha}
    - {\delta^\sigma}_\alpha\,\Metricc_{\beta\gamma}\r)
\end{equation}
where~$\Gcurv = (\rmg\tmg-\smg^2)/\wmg^4$ is the Gaussian curvature
and~$\Metricc_{\beta\alpha}$ is given by Eq.~\eqref{eq:GMonge}.
Hence, a simplified form of~\eqref{eq:dgeodesic} for surfaces is
\begin{equation}
  \frac{D^2}{D\time^2}\delta\xc^\sigma
  + \Gcurv\l(\langle\dot\xc,\dot\xc\rangle\,\delta\xc^\sigma
  - \langle\dot\xc,\delta\xc\rangle\,\dot\xc^\sigma\r) = 0
  \label{eq:dgeodesic2}
\end{equation}
where the inner product is defined by~$\langle V, W\rangle \ldef
\Metricc_{\alpha\beta}V^\alpha W^\beta$.

Note that the gravitational term does not enter
Eq.~\eqref{eq:dgeodesic} directly, though it does
through~\eqref{eq:geodesic}.  For the rest of this discussion we will
assume~$\grav=0$, since it simplifies the discussion considerably.  If
that is then case, then it is easy to show that
\begin{equation}
  \frac{D}{D\time}\langle\dot\xc,\delta\xc\rangle =
  \langle\dot\xc,\frac{D}{D\time}\delta\xc\rangle\,,\qquad
  \frac{D^2}{D\time^2}\langle\dot\xc,\delta\xc\rangle = 0
\end{equation}
which means that if we choose the initial~$\delta\xc^\alpha$ such
that~$\langle\dot\xc,\delta\xc\rangle =
\langle\dot\xc,{D}\delta\xc/{D\time}\rangle = 0$,
then~$\langle\dot\xc,\delta\xc\rangle$ remains zero for all time.
With this choice initial condition, the
geodesic deviation equation~\eqref{eq:dgeodesic2} finally takes the form
\begin{equation}
  \frac{D^2}{D\time^2}\delta\xc^\sigma
  + \Gcurv\langle\dot\xc,\dot\xc\rangle\,\delta\xc^\sigma = 0.
  \label{eq:dgeodesic3}
\end{equation}
Now we can ask under what condition the substrate shape will be
favourable to chaotic geodesics.  Since Eq.~\eqref{eq:dgeodesic3}
resembles an oscillator equation,
and~$\langle\dot\xc,\dot\xc\rangle\ge 0$, we see that negative Gaussian
curvature will favour divergence of trajectories.

\begin{figure}
\begin{center}
  \includegraphics[width=.6\textwidth]{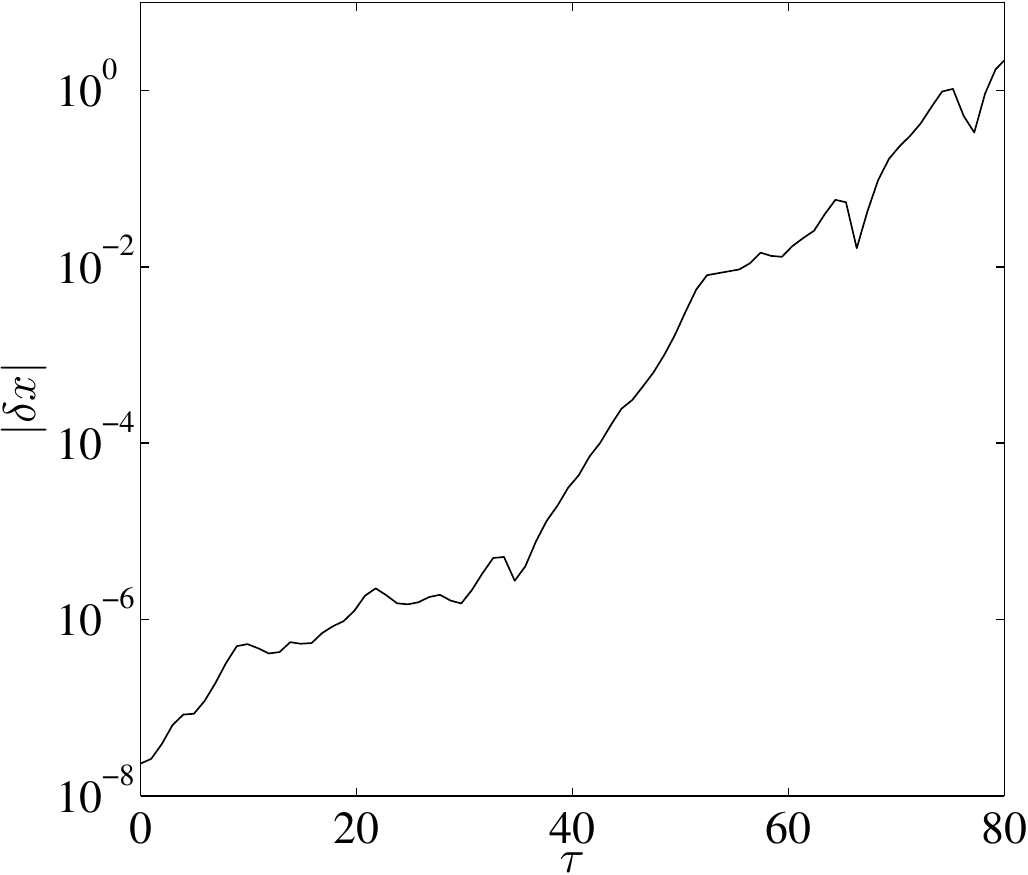}
\end{center}

\caption{The Cartesian distance~$|\delta\xc|$ between two trajectories, for
  the same parameters as in Fig.~\ref{fig:grav} with~$\fmg_0=0.5$.  The
  trajectories diverge extremely rapidly, consistent with chaotic behaviour.}
\label{fig:expsep}
\end{figure}

We have not yet solved Eq.~\eqref{eq:dgeodesic} for~$\delta\xc^\sigma$, but a
comparison of the distance between two initially very close trajectories is
shown in Fig.~\ref{fig:expsep}, for the same parameters as in
Fig.~\ref{fig:grav} ($\fmg_0=0.5$).
Unsurprisingly, the plot confirms exponential growth, demonstrating at least
numerically that chaos is indeed present.

\section{Discussion}
\label{sec:discussion}

We have shown that the flow of a shallow layer of inviscid,
irrotational fluid on a curved substrate leads to particle
trajectories that follow geodesics in the curved space, subject to
gravity.  We have displayed the range of behaviour that these
geodesics can exhibit, from regular to chaotic.

As Fig.~\ref{fig:cylinder} shows, the theory is not likely to be valid
much beyond the point where characteristics cross, and viscosity also
causes important corrections.  Another effect we ignored is the
possibility that centrifugal forces can cause the fluid to spin out
and detach from the substrate.~\cite{Edwards2008}  Experiments are
needed to determine to what extent the chaos observed here is
reproduced in reality.  If chaos is indeed prevalent, then perhaps
chaotic advection can be exploited in some applications to enhance
mixing in shallow layers.

We made the case in the introduction that chaos in the geodesic
equations was a subject worthy of study on its own.  The emergence of
chaotic behaviour as a function of Gaussian curvature, as embodied by
Eq.~\eqref{eq:dgeodesic2}, should be a rich subject of study, in
particular because of the simple form this equation takes on a
surface.

We note in closing that a similar study can be made for viscous thin
films.~\cite{ThiffeaultKamhawi2006} However, trajectories there are
much less prone to chaotic behaviour, because of the diminished role
of inertia.


\bibliography{journals_abbrev,articles,geodesic_deviation}

\begin{thebibliography}{20}%
\makeatletter
\providecommand \@ifxundefined [1]{%
 \@ifx{#1\undefined}
}%
\providecommand \@ifnum [1]{%
 \ifnum #1\expandafter \@firstoftwo
 \else \expandafter \@secondoftwo
 \fi
}%
\providecommand \@ifx [1]{%
 \ifx #1\expandafter \@firstoftwo
 \else \expandafter \@secondoftwo
 \fi
}%
\providecommand \natexlab [1]{#1}%
\providecommand \enquote  [1]{``#1''}%
\providecommand \bibnamefont  [1]{#1}%
\providecommand \bibfnamefont [1]{#1}%
\providecommand \citenamefont [1]{#1}%
\providecommand \href@noop [0]{\@secondoftwo}%
\providecommand \href [0]{\begingroup \@sanitize@url \@href}%
\providecommand \@href[1]{\@@startlink{#1}\@@href}%
\providecommand \@@href[1]{\endgroup#1\@@endlink}%
\providecommand \@sanitize@url [0]{\catcode `\\12\catcode `\$12\catcode
  `\&12\catcode `\#12\catcode `\^12\catcode `\_12\catcode `\%12\relax}%
\providecommand \@@startlink[1]{}%
\providecommand \@@endlink[0]{}%
\providecommand \url  [0]{\begingroup\@sanitize@url \@url }%
\providecommand \@url [1]{\endgroup\@href {#1}{\urlprefix }}%
\providecommand \urlprefix  [0]{URL }%
\providecommand \Eprint [0]{\href }%
\providecommand \doibase [0]{http://dx.doi.org/}%
\providecommand \selectlanguage [0]{\@gobble}%
\providecommand \bibinfo  [0]{\@secondoftwo}%
\providecommand \bibfield  [0]{\@secondoftwo}%
\providecommand \translation [1]{[#1]}%
\providecommand \BibitemOpen [0]{}%
\providecommand \bibitemStop [0]{}%
\providecommand \bibitemNoStop [0]{.\EOS\space}%
\providecommand \EOS [0]{\spacefactor3000\relax}%
\providecommand \BibitemShut  [1]{\csname bibitem#1\endcsname}%
\let\auto@bib@innerbib\@empty
\bibitem [{\citenamefont {Arnold}(1966)}]{Arnold1966a}%
  \BibitemOpen
  \bibfield  {author} {\bibinfo {author} {\bibfnamefont {V.~I.}\ \bibnamefont
  {Arnold}},\ }\href@noop {} {\bibfield  {journal} {\bibinfo  {journal} {Ann.
  Inst. Fourier}\ }\textbf {\bibinfo {volume} {16}},\ \bibinfo {pages} {319}
  (\bibinfo {year} {1966})}\BibitemShut {NoStop}%
\bibitem [{\citenamefont {Arnold}(1969)}]{Arnold1969b}%
  \BibitemOpen
  \bibfield  {author} {\bibinfo {author} {\bibfnamefont {V.~I.}\ \bibnamefont
  {Arnold}},\ }\href@noop {} {\bibfield  {journal} {\bibinfo  {journal} {Usp.
  Mat. Nauk.}\ }\textbf {\bibinfo {volume} {24}},\ \bibinfo {pages} {225}
  (\bibinfo {year} {1969})}\BibitemShut {NoStop}%
\bibitem [{\citenamefont {Arnold}(1989)}]{Arnold}%
  \BibitemOpen
  \bibfield  {author} {\bibinfo {author} {\bibfnamefont {V.~I.}\ \bibnamefont
  {Arnold}},\ }\href@noop {} {\emph {\bibinfo {title} {Mathematical Methods of
  Classical Mechanics}}},\ \bibinfo {edition} {2nd}\ ed.\ (\bibinfo
  {publisher} {Springer-Verlag},\ \bibinfo {address} {New York},\ \bibinfo
  {year} {1989})\BibitemShut {NoStop}%
\bibitem [{\citenamefont {Marsden}\ and\ \citenamefont
  {Ratiu}(1994)}]{MarsdenRatiu}%
  \BibitemOpen
  \bibfield  {author} {\bibinfo {author} {\bibfnamefont {J.~E.}\ \bibnamefont
  {Marsden}}\ and\ \bibinfo {author} {\bibfnamefont {T.~S.}\ \bibnamefont
  {Ratiu}},\ }\href@noop {} {\emph {\bibinfo {title} {Introduction to Mechanics
  and Symmetry}}}\ (\bibinfo  {publisher} {Springer-Verlag},\ \bibinfo
  {address} {Berlin},\ \bibinfo {year} {1994})\BibitemShut {NoStop}%
\bibitem [{\citenamefont {Arnold}\ and\ \citenamefont
  {Khesin}(1998)}]{ArnoldTopo}%
  \BibitemOpen
  \bibfield  {author} {\bibinfo {author} {\bibfnamefont {V.~I.}\ \bibnamefont
  {Arnold}}\ and\ \bibinfo {author} {\bibfnamefont {B.~A.}\ \bibnamefont
  {Khesin}},\ }\href@noop {} {\emph {\bibinfo {title} {Topological Methods in
  Hydrodynamics}}}\ (\bibinfo  {publisher} {Springer-Verlag},\ \bibinfo
  {address} {New York},\ \bibinfo {year} {1998})\BibitemShut {NoStop}%
\bibitem [{\citenamefont {Watanabe}(2007)}]{Watanabe2007}%
  \BibitemOpen
  \bibfield  {author} {\bibinfo {author} {\bibfnamefont {Y.}~\bibnamefont
  {Watanabe}},\ }\href@noop {} {\bibfield  {journal} {\bibinfo  {journal}
  {Physica D}\ }\textbf {\bibinfo {volume} {225}},\ \bibinfo {pages} {197}
  (\bibinfo {year} {2007})}\BibitemShut {NoStop}%
\bibitem [{\citenamefont {Wald}(1984)}]{Wald}%
  \BibitemOpen
  \bibfield  {author} {\bibinfo {author} {\bibfnamefont {R.~M.}\ \bibnamefont
  {Wald}},\ }\href@noop {} {\emph {\bibinfo {title} {General Relativity}}}\
  (\bibinfo  {publisher} {University of Chicago Press},\ \bibinfo {address}
  {Chicago},\ \bibinfo {year} {1984})\BibitemShut {NoStop}%
\bibitem [{\citenamefont {Rienstra}(1996)}]{Rienstra1996}%
  \BibitemOpen
  \bibfield  {author} {\bibinfo {author} {\bibfnamefont {S.~W.}\ \bibnamefont
  {Rienstra}},\ }\href@noop {} {\bibfield  {journal} {\bibinfo  {journal} {Z.
  Angew. Math. Mech.}\ }\textbf {\bibinfo {volume} {76}},\ \bibinfo {pages}
  {423} (\bibinfo {year} {1996})}\BibitemShut {NoStop}%
\bibitem [{\citenamefont {Edwards}\ \emph {et~al.}(2008)\citenamefont
  {Edwards}, \citenamefont {Howison}, \citenamefont {Ockendon},\ and\
  \citenamefont {Ockendon}}]{Edwards2008}%
  \BibitemOpen
  \bibfield  {author} {\bibinfo {author} {\bibfnamefont {C.~M.}\ \bibnamefont
  {Edwards}}, \bibinfo {author} {\bibfnamefont {S.~D.}\ \bibnamefont
  {Howison}}, \bibinfo {author} {\bibfnamefont {H.}~\bibnamefont {Ockendon}}, \
  and\ \bibinfo {author} {\bibfnamefont {J.~R.}\ \bibnamefont {Ockendon}},\
  }\href {\doibase 10.1093/imamat/hxm064} {\bibfield  {journal} {\bibinfo
  {journal} {IMA J. Appl. Math.}\ }\textbf {\bibinfo {volume} {73}},\ \bibinfo
  {pages} {137} (\bibinfo {year} {2008})}\BibitemShut {NoStop}%
\bibitem [{\citenamefont {Bouchut}(1994)}]{Bouchut1994}%
  \BibitemOpen
  \bibfield  {author} {\bibinfo {author} {\bibfnamefont {F.}~\bibnamefont
  {Bouchut}},\ }in\ \href@noop {} {\emph {\bibinfo {booktitle} {Advances in
  Kinetic Theory and Computing}}},\ \bibinfo {series} {Advances in Mathematics
  for Applied Sciences}, Vol.~\bibinfo {volume} {22},\ \bibinfo {editor}
  {edited by\ \bibinfo {editor} {\bibfnamefont {B.}~\bibnamefont {Perthame}}}\
  (\bibinfo  {publisher} {World Scientific},\ \bibinfo {year}
  {1994})\BibitemShut {NoStop}%
\bibitem [{\citenamefont {Li}\ \emph {et~al.}(1998)\citenamefont {Li},
  \citenamefont {Zhang},\ and\ \citenamefont {Yang}}]{LiZhangYang}%
  \BibitemOpen
  \bibfield  {author} {\bibinfo {author} {\bibfnamefont {J.}~\bibnamefont
  {Li}}, \bibinfo {author} {\bibfnamefont {T.}~\bibnamefont {Zhang}}, \ and\
  \bibinfo {author} {\bibfnamefont {S.}~\bibnamefont {Yang}},\ }\href@noop {}
  {\emph {\bibinfo {title} {Two-dimensional {R}iemann Problems in Gas
  Dynamics}}}\ (\bibinfo  {publisher} {Chapman \& Hall/{CRC} Press},\ \bibinfo
  {address} {Boca Raton, {FL}},\ \bibinfo {year} {1998})\BibitemShut {NoStop}%
\bibitem [{\citenamefont {Yang}(1999)}]{Yang1999}%
  \BibitemOpen
  \bibfield  {author} {\bibinfo {author} {\bibfnamefont {H.}~\bibnamefont
  {Yang}},\ }\href@noop {} {\bibfield  {journal} {\bibinfo  {journal} {J. Diff.
  Eqns.}\ }\textbf {\bibinfo {volume} {159}},\ \bibinfo {pages} {447} (\bibinfo
  {year} {1999})}\BibitemShut {NoStop}%
\bibitem [{\citenamefont {Li}(2001)}]{Li2001}%
  \BibitemOpen
  \bibfield  {author} {\bibinfo {author} {\bibfnamefont {J.}~\bibnamefont
  {Li}},\ }\href@noop {} {\bibfield  {journal} {\bibinfo  {journal} {Appl.
  Math. Lett.}\ }\textbf {\bibinfo {volume} {14}},\ \bibinfo {pages} {519}
  (\bibinfo {year} {2001})}\BibitemShut {NoStop}%
\bibitem [{\citenamefont {Kreyszig}(1959)}]{Kreyszig}%
  \BibitemOpen
  \bibfield  {author} {\bibinfo {author} {\bibfnamefont {I.}~\bibnamefont
  {Kreyszig}},\ }\href@noop {} {\emph {\bibinfo {title} {Differential
  Geometry}}}\ (\bibinfo  {publisher} {University of Toronto Press},\ \bibinfo
  {address} {Toronto},\ \bibinfo {year} {1959})\BibitemShut {NoStop}%
\bibitem [{\citenamefont {Flanders}(1990)}]{Flanders}%
  \BibitemOpen
  \bibfield  {author} {\bibinfo {author} {\bibfnamefont {H.}~\bibnamefont
  {Flanders}},\ }\href@noop {} {\emph {\bibinfo {title} {Differential Forms
  with Applications to the Physical Sciences}}}\ (\bibinfo  {publisher}
  {Dover},\ \bibinfo {address} {New York},\ \bibinfo {year} {1990})\BibitemShut
  {NoStop}%
\bibitem [{\citenamefont {Synge}\ and\ \citenamefont {Schild}(1978)}]{Synge}%
  \BibitemOpen
  \bibfield  {author} {\bibinfo {author} {\bibfnamefont {J.~L.}\ \bibnamefont
  {Synge}}\ and\ \bibinfo {author} {\bibfnamefont {A.}~\bibnamefont {Schild}},\
  }\href@noop {} {\emph {\bibinfo {title} {Tensor Calculus}}}\ (\bibinfo
  {publisher} {Dover},\ \bibinfo {address} {New York},\ \bibinfo {year}
  {1978})\BibitemShut {NoStop}%
\bibitem [{\citenamefont {Schutz}(1980)}]{Schutz}%
  \BibitemOpen
  \bibfield  {author} {\bibinfo {author} {\bibfnamefont {B.}~\bibnamefont
  {Schutz}},\ }\href@noop {} {\emph {\bibinfo {title} {Differential
  Geometry}}}\ (\bibinfo  {publisher} {Cambridge University Press},\ \bibinfo
  {address} {Cambridge, U.K.},\ \bibinfo {year} {1980})\BibitemShut {NoStop}%
\bibitem [{\citenamefont {Thiffeault}(2001)}]{Thiffeault2001e}%
  \BibitemOpen
  \bibfield  {author} {\bibinfo {author} {\bibfnamefont {J.-L.}\ \bibnamefont
  {Thiffeault}},\ }\href@noop {} {\bibfield  {journal} {\bibinfo  {journal} {J.
  Phys. A}\ }\textbf {\bibinfo {volume} {34}},\ \bibinfo {pages} {5875}
  (\bibinfo {year} {2001})}\BibitemShut {NoStop}%
\bibitem [{geo()}]{geodesic_deviation}%
  \BibitemOpen
  \href@noop {} {}\bibinfo {note} {For a step-by-step derivation of the
  geodesic deviation equation see
  \url{http://ion.uwinnipeg.ca/~vincent/4500.6-001/Cosmology/GeodesicDeviation%
.htm}.}\BibitemShut {Stop}%
\bibitem [{\citenamefont {Thiffeault}\ and\ \citenamefont
  {Kamhawi}(2006)}]{ThiffeaultKamhawi2006}%
  \BibitemOpen
  \bibfield  {author} {\bibinfo {author} {\bibfnamefont {J.-L.}\ \bibnamefont
  {Thiffeault}}\ and\ \bibinfo {author} {\bibfnamefont {K.}~\bibnamefont
  {Kamhawi}},\ }\href@noop {} {\enquote {\bibinfo {title} {Transport in thin
  gravity-driven flow over a curved substrate},}\ } (\bibinfo {year} {2006}),\
  \Eprint {http://arxiv.org/abs/arXiv:nlin/0607075} {arXiv:nlin/0607075}
  \BibitemShut {NoStop}%
\end{thebibliography}%

\end{document}